\documentclass[prb,twocolumn,superscriptaddress,showpacs]{revtex4-1}
\usepackage{graphicx}

\usepackage[colorlinks=true,citecolor=blue,urlcolor=blue]{hyperref}

\begin{document}

\title{Variational Monte Carlo simulations using tensor-product projected states}

\author{Olga Sikora}
\address{Department of Physics, National Taiwan University, Taipei 10607, Taiwan}
\author{Hsueh-Wen Chang}
\address{Department of Physics, National Taiwan University, Taipei 10607, Taiwan}
\author{Chung-Pin Chou}\email{cpc63078@gmail.com}
\address{Beijing Computational Science Research Center, Beijing 100084, China}
\author{Frank Pollmann}\email{frankp@pks.mpg.de}
\address{Max-Planck-Institut f{\"u}r Physik komplexer Systeme, 01187 Dresden, Germany}
\author{Ying-Jer Kao}\email{yjkao@phys.ntu.edu.tw}
\address{Department of Physics, National Taiwan University, Taipei 10607, Taiwan}

\begin{abstract}
We propose an efficient numerical method, which combines the
advantages of recently developed tensor-network based methods and
standard trial wave functions, to study the ground state
properties of quantum many-body systems.
In this approach, we apply a projector in the form of a tensor-product operator to an
input wave function, such as a Jastrow-type or Hartree-Fock wave
function, and optimize the tensor elements via variational Monte
Carlo.
The entanglement already contained in the input wave function can
considerably reduce the bond dimensions compared to the regular tensor-product state representation.
In particular, this allows us to also represent states that do not
obey the area law of entanglement entropy.
In addition, for fermionic systems, the fermion sign structure can be encoded in the input wave
function.
We show that the optimized states provide good approximations of the
ground-state energy and correlation functions in the cases of
two-dimensional bosonic and fermonic systems.
\end{abstract}

\pacs{
71.10.Fd,   
02.70.Ss,   
05.10.Cc  
}
\maketitle

%%%%%%%%%%%%%%%%%%%%%%%%%%%%%%%%%%%%%%%%%%%%%%%%%%%%%
\section{Introduction}\label{sec1}
%%%%%%%%%%%%%%%%%%%%%%%%%%%%%%%%%%%%%%%%%%%%%%%%%%%%%
Interacting electron and spin models in two dimensions (2D) exhibit
some of the most interesting phenomena in condensed-matter physics,
e.g., superconductivity and topologically ordered spin-liquid phases.
\cite{XGWenBOOK,Balents10}
The simulation of these quantum many-body systems is, however, a big
challenge for all available computation methods.
For example, exact diagonalization (ED) allows only for very limited
system sizes, the density matrix renormalization group (DMRG)
\cite{WhitePRL92} is extremely efficient in one-dimensional (1D)
systems but becomes very challenging in higher dimensional systems,
and quantum Monte Carlo (QMC) methods \cite{LindenPR92} suffer from
the infamous sign problem in fermionic and frustrated systems.

Over the past years, the class of tensor-product states has been shown to be
a very promising tool.
Most successful are matrix product states (MPS),
\cite{OstlundPRL95} which are the underlying base of the DMRG
method, providing the most powerful tool to study 1D systems.
Various generalization of the MPS approach to 2D have been proposed,
including rather direct generalizations such as  projected entangled pair states or tensor product states (TPS) \cite{Verstraete04,VerstraetePRA04,ShiPRA06,SchuchPRL08, SandvikPRL08, MurgPRB10} as well as  correlator
product states or entangled-plaquette states.\cite{MezzacapoNJP09,ChanglaniPRB09,Hubener09-11,HubenerNJP10}
TPS-based algorithms have been successfully applied
to various frustrated quantum spin systems and hard-core bosons in 2D
\cite{MurgPRA07,MurgPRB09,BauerJSM09,Wang11}.
Based on the concept of entanglement renormalization, the so-called
tensor renormalization group (TRG) and multi scale entanglement renormalization group (MERA) methods have
been developed and successfully applied to lattice systems.\cite{VidalPRL07,GuPRB08,JiangPRL08,YJK11-12,GuPRB13}
Yet, due to the scaling of computational complexity as a function of
the dimension of the tensors (i.e., the bond dimension), the
applicability of the algorithms is still limited.
This limitation motivates a search for alternative
tensor-network-based algorithms.
Recently, variational Monte Carlo (VMC) methods have been proposed,
improving a tensor-product state (TPS) by stochastically applying
projectors which filter out excited states.\cite{Wouters14,
Clark14} 
In an approach closely related to our work,  wave functions combining a correlator
 product state and a Slater determinant (or a Pfaffian) were studied. \cite{Neu84, Neu85}
Hartree-Fock and configurational interaction based input wave functions 
were also used to achieve  better accuracy and lower the
required bond dimension in the DMRG procedure. \cite{LegPRB03, LegPRA11}
These results suggest that it is important to choose the basis wave functions
that already represent some key features of the physics of the system under
investigation.

In this paper, we propose a method which combines 2D tensor product
 states with standard trial wave functions.
Our study is a generalization of a recent work by Chou
\textit{et al.} \cite{Chou11} in which matrix-product based projected
wave functions were considered as trial wave functions for VMC
simulations in a 1D model.
The main idea is that we can use physical intuition to choose an
input wave function that already contains some of the features of the
ground state which are difficult to be captured by a TPS with small
bond dimension.
This way we can reduce the  computational costs arising from the
tensor contraction and still have a very good approximation of the
ground-state wave function.
Furthermore, using the VMC method to sample energies, we ensure
that we  approach the ground-state energy from above.
We demonstrate the effectiveness of our approach by benchmarking the
ground-state energies and correlations for several 2D interacting
quantum systems, such as spinless fermionic and hardcore bosonic
$t$-$V$ models, as well as the spinful fermionic Hubbard model.
In these quantum models, we find that a suitable Jastrow-type or
Hartree-Fock wave function greatly helps the TPS to approach the
ground state.

This paper is organized as follows.
We begin by introducing the main concept of our approach and review
some relevant TPS based methods in Sec.~\ref{sec2}.
In Sec.~\ref{sec3}, we present our results and benchmarks for
different bosonic and fermionic models.
We conclude with a summary and discussion in Sec.~\ref{sec4}.

%%%%%%%%%%%%%%%%%%%%%%%%%%%%%%%%%%%%%%%%%%%%%%%%%%%%%
\section{Methods}\label{sec2}
%%%%%%%%%%%%%%%%%%%%%%%%%%%%%%%%%%%%%%%%%%%%%%%%%%%%%

%++++++++++++++++++++++++++++++++++++++++++++++++++++++++++++++++%
\subsection{Tensor-network projected ansatz}\label{tnpa}
\begin{figure}[t]
\begin{center}
\includegraphics[width=0.45\textwidth]{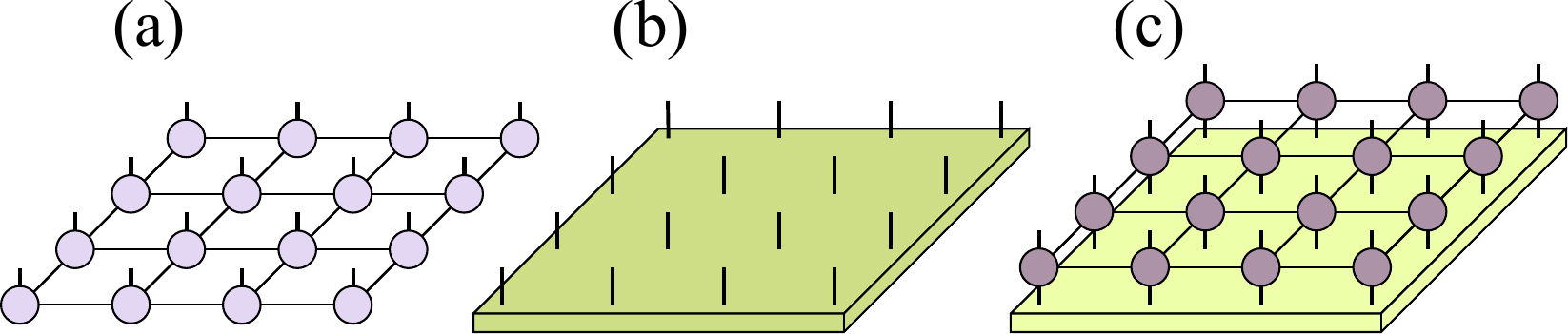}
\end{center}
\caption{Representations of (a) a TPS whose closed links represent
bonds of dimension $D$ and open links physical indices; (b) an input
$N$-particle wave function $|\Psi_{0}\rangle$ with indices depicted
as open links; (c) a tensor-product projection operator acting on
the wave function $|\Psi_{0}\rangle$ highlighted in green. }
\label{fig:tpps}
\end{figure}

%++++++++++++++++++++++++++++++++++++++++++++++++++++++++++++++++%
We consider tensor-network projected wave functions as a class of
quantum states for Monte Carlo optimization, and  in this section we
discuss a general formula for such states.
First, a TPS can be defined as
\begin{eqnarray}
|\Psi_{\mathrm{TPS}}\rangle=\sum_{j_1,\ldots,j_N}\mathrm{tTr}\left(\prod_{i=1}^{N}\hat{T}_{[i]}^{j_{i}}\right)|\alpha\rangle,\label{eq:tps}
\end{eqnarray}
where $\alpha\equiv\{j_{1},j_{2},\cdots,j_{N}\}$ represents a
many-body configuration in the lattice system of size $N$,
$\hat{T}_{[i]}^{j_{i}}$ is the rank-4 tensor with bond dimension $D$,
and the physical index $j_i$ representing a quantum state at site
$i$.
The configuration weight is given by contraction of a network of
tensors as shown in Fig.~\ref{fig:tpps}(a), denoted as a tensorial
trace $\mathrm{tTr}$ over all virtual bond indices.
In this work we use this representation as one of the trial wave
functions for simulating the bosonic system.
%
%------------------------------------------------------------------------------------%
%------------------------------------------------------------------------------------%

More generally, a representation of a quantum many-body state, which
we call the tensor-product projected state (TPPS), can be written in
terms of the projection operators acting on an input wave function
$|\Psi_{0}\rangle$:
\begin{eqnarray}
|\Psi_{\mathrm{TPPS}}\rangle=\sum_{j_1,\ldots,j_N}\mathrm{tTr}\left(\prod_{i=1}^{N}P^{j_{i}}_{[i]}\right)|\Psi_{0}\rangle,\label{eq:tps1}
\end{eqnarray}
where $P^{j_{i}}_{[i]}=\hat{T}_{[i]}^{j_{i}}|j_{i}\rangle\langle
j_{i}|$. If we assume
$|\Psi_{0}\rangle=\sum_{\alpha}|\alpha\rangle$, i.e., a wave function
with equal probability distribution of all configurations (a
site-factorized state), we recover the TPS wave function of
Eq.~(\ref{eq:tps}).
However, we can also choose a different trial state, which might be
closer to the ground state of a given system (e.g., incorporate entanglement)
and thus require a smaller bond dimension
of the tensor network.\cite{Chou11}
A schematic representation of the TPPS wave function is shown in
Fig.~\ref{fig:tpps}(c).
Expressed in a configuration basis, the TPPS is given by
\begin{eqnarray}
|\Psi_{\mathrm{TPPS}}\rangle=\sum_{j_1,\ldots,j_N}\mathrm{tTr}\left(\prod_{i=1}^{N}\hat{T}_{[i]}^{j_{i}}\right)\langle\alpha|\Psi_{0}\rangle|\alpha\rangle.\label{eq:equ1}
\end{eqnarray}
In this work, we examine several different input states
$|\Psi_{0}\rangle$: (1) Jastrow wave function for hardcore bosons;
(2) ground state  wave function for noninteracting spinless fermions, i.e. Slater determinant (SL); (3) SL,
$d$-wave BCS (d-BCS) and spin-density-wave (SDW) states for spinful
fermions.

%++++++++++++++++++++++++++++++++++++++++++++++++++++++++++++++++%
\subsection{Variational optimization of the wave function}
%++++++++++++++++++++++++++++++++++++++++++++++++++++++++++++++++%
%
To optimize the ground-state energy, we apply standard quantum VMC methods \cite{Sorella01} to
the TPPS wave function defined in Eq.~(\ref{eq:equ1}).
The expectation values of the observables are obtained as the
average over the values for configurations visited during a Markov
walk, and the probability distribution is given by coefficients of
the wave function.
The weight of the wave function for a given configuration $\alpha$
is
\begin{equation}
W(\alpha)={\rm
tTr}\left(\prod_{i=1}^{N}\hat{T}_{[i]}^{j_{i}}\right)\langle\alpha|\Psi_{0}\rangle,
\label{ws}
\end{equation}
where the tensorial trace corresponds to the contraction of a
single-layer tensor network with $N$ open indices given by $\alpha$.
The energy, for given tensors $\hat{T}_{[i]}^{j_{i}}$, can be
written in the form appropriate for Monte Carlo sampling,
\begin{eqnarray}
E&=&\langle E(\alpha)\rangle=\frac{1}{Z}\sum_\alpha
W^2(\alpha)E(\alpha),\\
\quad Z&=&\sum_\alpha W^2(\alpha),\label{esdetail}
\end{eqnarray}
where the estimator is given by
\begin{equation}
E(\alpha)=\sum_{\alpha'} \frac{W(\alpha')}{W(\alpha)}\langle
\alpha'|\hat H|\alpha\rangle.\label{es}
\end{equation}
The energy can be evaluated using importance sampling of the
configurations according to the weight $W^2(\alpha)$.

We employ the stochastic reconfiguration (SR) method
\cite{Sorella01} to handle the multi-variable optimization of a wave
function in the part of the Hilbert space spanned by its degrees of
freedom.
The SR method is a projector based optimization using the Hamiltonian
to filter out the lowest energy state within the available subspace
of the Hilbert space.
The trial wave function can be also expanded in the variational
parameters, which leads to a system of linear equations, with a new set
of parameters as solutions.
For the TPPS wave function, both the elements of tensors and
parameters of the trial wave function might be variationally
optimized.
More details about the SR method can be found in
Refs.~\onlinecite{YunokiPRB06,Capello06}. 
The Monte Carlo sampling scheme ensures that the wave function is variational, and the energy fulfills the variational principle, i.e., the approximated energy provides an upper bound for the exact energy.
 In contrast, this is not necessarily the case when the wave function is obtained using the simplified update and the tensor network is contracted approximately.\cite{JiangPRL08}

%++++++++++++++++++++++++++++++++++++++++++++++++++++++++++++++++%
\subsection{Contraction of the tensor network}
\begin{figure}[t]
\begin{center}
\includegraphics[width=8cm]{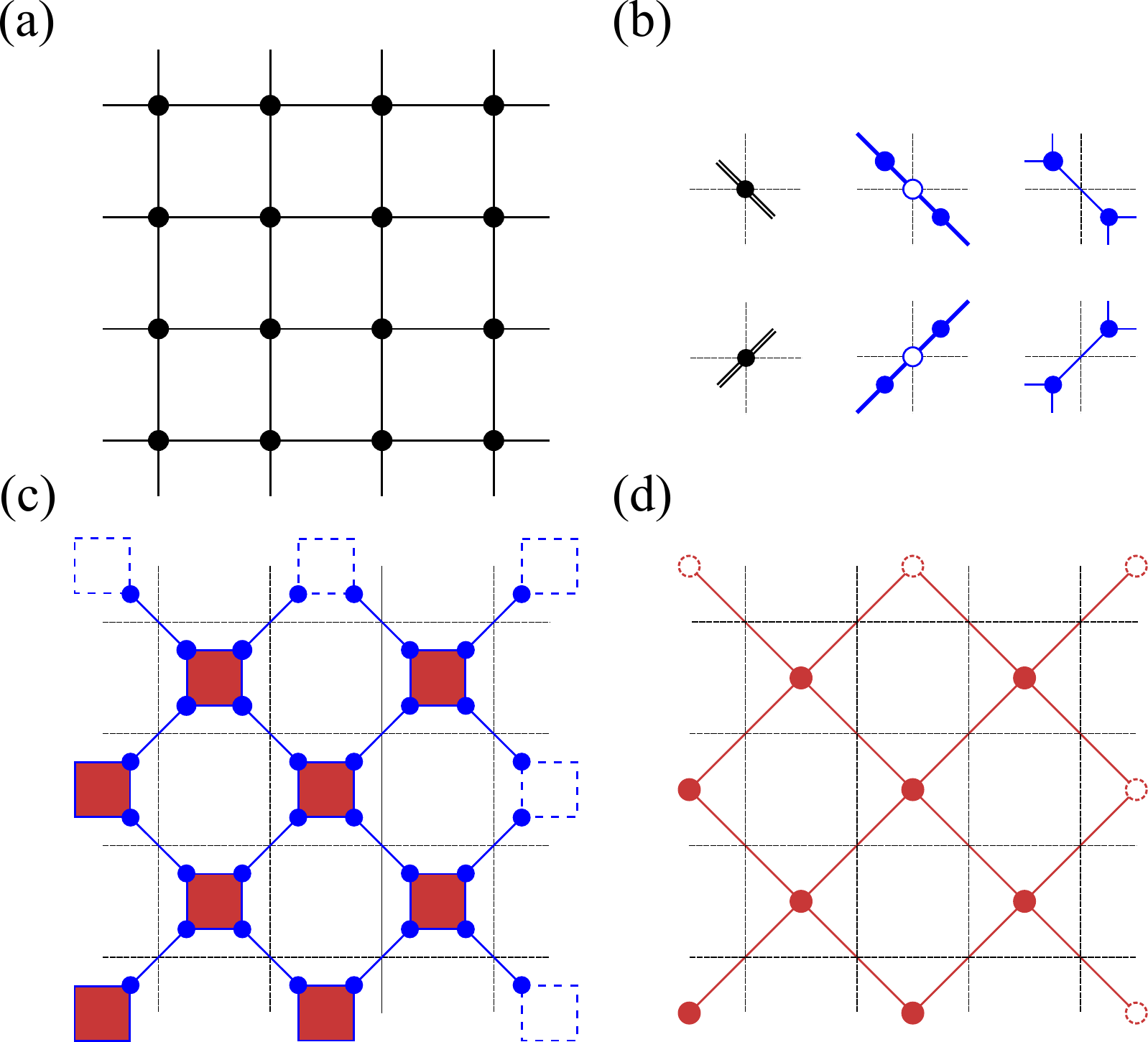}
\end{center}
\caption{TRG algorithm: (a) Bipartite network of sixteen tensors with periodic
boundary conditions, where the tensor elements depend on the value of the
physical index  (site occupation) and the sublattice index. (b) Singular
value decomposition and truncation at $D_f$,
the middle matrix represented by empty circle is then absorbed into
tensors (filled circles). (c) Contracting the tensors in the centers of the plaquettes yields a network of eight tensors (d).}
\label{fig:trg_scheme}
\end{figure}

%++++++++++++++++++++++++++++++++++++++++++++++++++++++++++++++++%
During the simulation, we need to evaluate the contribution from the
tensor network to the weight.
We perform exact tensor contraction for $4\times4$ systems with bond
dimension $D=2$ in the $t$-$V$ model and for all bond dimensions
considered in the study of the Hubbard model.
To efficiently evaluate the tensor network weight beyond this system
size and/or bond dimension, we adopt the coarse-graining TRG method
tailored for the Monte Carlo sampling.
\cite{GuPRB08,JiangPRL08,Wang11}
We briefly review the TRG procedure for contracting a $2^n\times2^n$
tensor network, and illustrate it schematically in Fig.~\ref{fig:trg_scheme}(a)-(d)
for a $4\times4$ system.
The tensor contraction is done by coarse-graining the lattice until
the $2\times2$ tensor network can be directly contracted.
In each step, we perform a singular value decomposition (SVD) to
decompose a rank-4 tensor on each site of the lattice into two
rank-3 tensors, keeping only the largest $D_f$ singular values, as
shown in Fig.~\ref{fig:trg_scheme}(b).
Figures~\ref{fig:trg_scheme}(c) and (d) show how the tensors are
contracted to form a smaller lattice, for which the
procedure is repeated. 
In the current simulation, the convergence can be reached by setting the cutoff 
$D_{f}$ to very small values of $2$ or $3$, retaining relatively low
computational costs.
In general, however, the cost of contracting a tensor network with a given cutoff $D_f$ 
grows rapidly with $D$, e.g. it takes almost 20 times longer to contract a $D$=3 than a $D$=2 
TPS with $D_f=8$.  
It is thus important to start with a good trial wave function rather than to blindly
increase the bond dimension of the tensors. \\
During the simulation, we apply local updates and change
the physical state accordingly.
We can lower the computational cost of tensor contraction by using
part of the results from the previous step and updating only the
elements that are changed at each stage of the coarse graining
procedure.\cite{Wang11}
This way of updating elements is particularly helpful for larger
lattice systems.
In the following section, we illustrate the concept of the TPPS wave
function using several examples.

%------------------------------------------------------------------------------------%
%------------------------------------------------------------------------------------%

%%%%%%%%%%%%%%%%%%%%%%%%%%%%%%%%%%%%%%%%%%%%%%%%%%%%%
\section{Numerical Results}\label{sec3}
%%%%%%%%%%%%%%%%%%%%%%%%%%%%%%%%%%%%%%%%%%%%%%%%%%%%%

%++++++++++++++++++++++++++++++++++++++++++++++++++++++++++++++++%
\subsection{Hardcore bosons}
%++++++++++++++++++++++++++++++++++++++++++++++++++++++++++++++++%

%------------------------------------------------------------------------------------%
\begin{figure*}[t]
\begin{center}
\includegraphics[width=\textwidth]{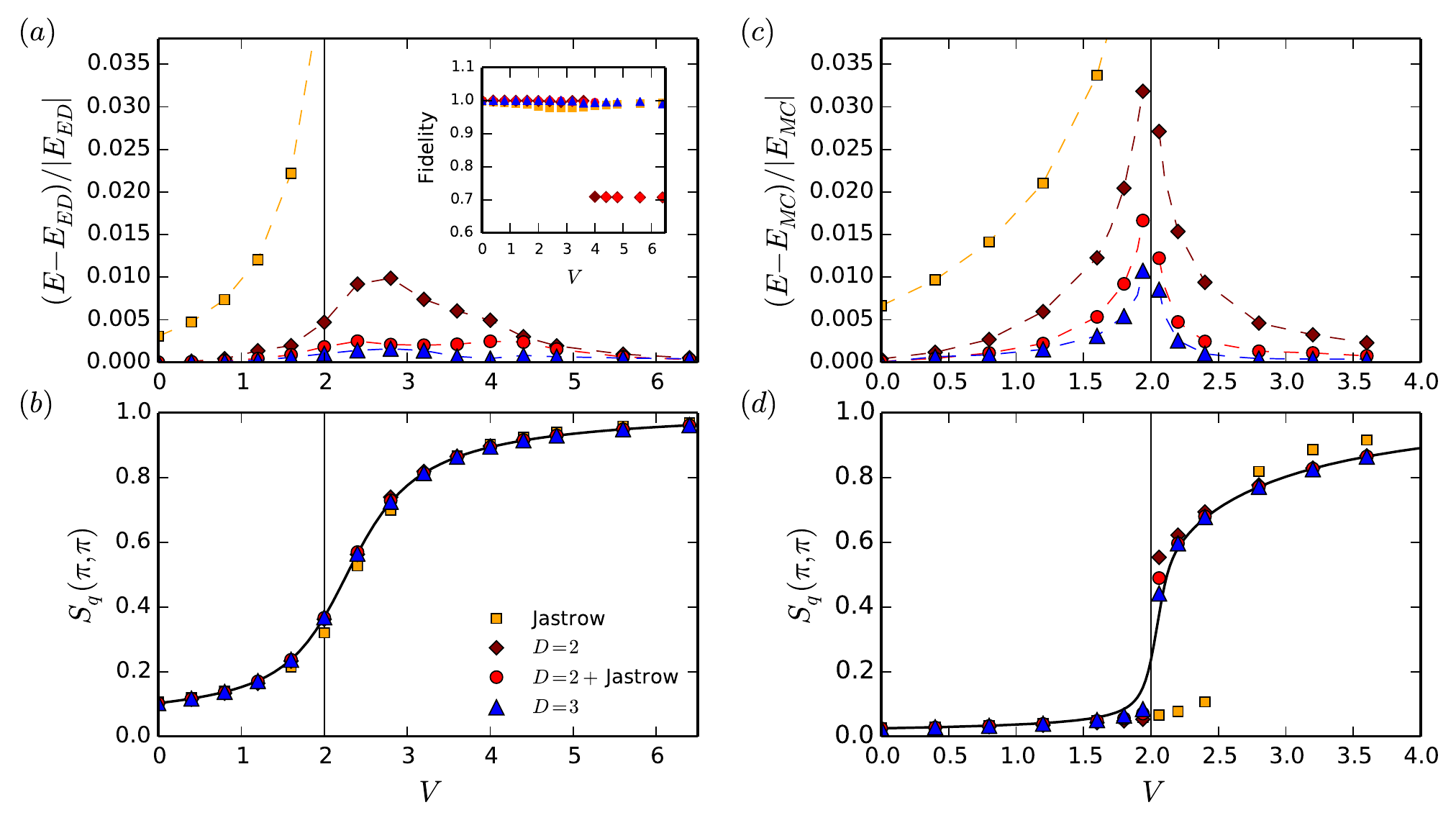}
\end{center}
\caption{Results of VMC simulations of hardcore bosons on $4\times4$
(left panel) and $8\times8$ (right panel) lattice with periodic
boundary conditions obtained for different input wave functions: a
two-parameter Jastrow wave function (orange squares), a TPS with the
bond dimension $D=2$ (maroon diamonds), a projected $D=2$ TPS on the
Jastrow factor wave function (red circles), and a TPS with $D=3$
(blue triangles). The upper panels show a relative energy error
compared to (a) the ED and (c) the QMC simulations, and the inset shows 
fidelity obtained for the $4\times4$ system. The lower panels
display the structure factor $S_q(\pi,\pi)$: (b) the ED and (d) the
QMC results are denoted by solid lines. } \label{fig:bs}
\end{figure*}
%------------------------------------------------------------------------------------%

We first discuss the hardcore boson model with  nearest-neighbor
repulsion (the $t$-$V$ model) at half filling on the square lattice.
The motivation for this study is to test if the TPPS approach with suitable
input wave functions can significantly improve the
ground-state energy and correlations compared to TPS.

The Hamiltonian of this model is given by
\begin{equation}
H_{tV}=-t\sum_{\langle ij\rangle}(b_i^{\dagger}b_j+ H.c.)+V\sum_{\langle ij\rangle}n_i n_j,\label{eq:ham}
\end{equation}
where $t=1$ is the hopping integral, $V$ denotes the repulsion
strength, $b_i^{(\dag)}$ annihilates (creates) a hardcore boson on site $i$, and $n_i=b_i^{\dagger}b_i$.
The hardcore boson
can be mapped to the spin-1/2 antiferromagnetic (AF) XXZ
model, with half filling corresponding to the zero magnetization
sector.\cite{Hebert01}
The  hopping terms translate to the spin flipping, while
the interaction terms map to the Ising interactions.
The point $V=2t$ corresponds to the Heisenberg point
of the XXZ model, and the XY-Ising change in universality class
corresponds to the transition between a superfluid
phase and a charge-density-wave (CDW) insulator at $V=2t$.

We simulate the model Hamiltonian Eq.~(\ref{eq:ham}) by examining several input wave
functions including the Jastrow-type wave function, the TPS and the
TPPS with a Jastrow input wave function.
The simplest variational wave function we consider for this model is
a Jastrow-type wave function $|\Psi_{\mathrm{Jastrow}}\rangle$ with
two variational parameters ($\gamma$ and $\beta$) associated with
the local particle-particle correlations.
The weight of a given configuration $\alpha$ is given by
\begin{equation}
\langle\alpha|\Psi_{\mathrm
{Jastrow}}\rangle=\langle\alpha|\hat{J}_{\gamma,\beta}|\alpha\rangle.\label{eq:jastrow}
\end{equation}
Here the Jastrow operator is defined as
$\hat{J}_{\gamma,\beta}\equiv \exp({\gamma\sum_{\langle ij\rangle}n_i
n_j+\beta\sum_{\langle\langle ij\rangle\rangle}n_i n_j})$, where the
sum over $\langle ij\rangle$ counts the number of the
nearest-neighbor occupied sites and $\langle\langle ij\rangle
\rangle$ the number of the next-nearest-neighbor occupied sites.
Simulating the Jastrow-type wave function we find that the energy
error obtained with respect to the ED results grows rapidly with the
value of $V$ and reaches the value of about 20\% in the CDW
insulating phase shown in Fig.~\ref{fig:bs}(a).

For the TPS  [Eq.~(\ref{eq:tps})], we assume periodic boundary
conditions and a bipartite structure for tensor elements: all
tensors with the same physical index on a given sublattice ($A$ or
$B$) have the same elements, moreover, $\hat{T}_{[i\in
A]}^{1(0)}=\hat{T}_{[i\in B]}^{0(1)}$, where the upper (physical)
index denotes site occupation.
We further assume that the tensor elements are real and symmetric,
and that they have rotational symmetry, i.e., the tensor elements
are related,
\begin{equation}
\hat{T}_{ijkl}=\hat{T}_{jkli}=\hat{T}_{klij}=\hat{T}_{lijk},\label{eq:symmetry}
\end{equation}
where $i,j,k,l=1,\ldots,D$ are the indices for the virtual bonds.
We begin by optimizing the TPS with the bond dimension $D=2$.
One can see in Fig.~\ref{fig:bs}(a) that indeed the TPS is a much
better trial wave function than the simple Jastrow-type wave
function for all values of interaction $V$.
However, the computational cost to obtain reliable results from the
TPS is much higher even for a $4\times4$ lattice system and the
smallest nontrivial bond dimension $D=2$.
As we can expect, the TPS with $D=3$ offers much better description
to the ground state, especially around the transition to the CDW
insulator.
Increasing $D$ is a systematic way of improving the quality of the wave
function, but the contraction cost increases even further.
Moreover, the number of different tensor parameters that need to be
optimized increases (in our case from 12 to 48).

We now compare the TPS results to our TPPS [Eq.~(\ref{eq:equ1})]
approach in which we use a  two-parameter Jastrow input wave
function.
Including the two Jastrow parameters in the TPPS practically does
not increase the computational cost with respect to the TPS with $D=2$.
Fig.~\ref{fig:bs}(a) shows that the results in a $4\times4$ lattice
system look very promising in terms of the energy errors, which
are much closer to the results of the TPS with $D=3$, especially
where the TPS gives the maximal energy error for $V\approx3$.\\

In the inset to Fig.~\ref{fig:bs}(a) we show the overlap (fidelity) $F$ between 
the ED wave function $ |\Phi_{ED}\rangle$  and the wave functions $|\Psi \rangle$ 
optimized by our simulations for the $4\times4$ system.
\begin{equation}
F=|\langle\Psi | \Phi_{ED}\rangle|.
\label{eq:fidelity}
\end{equation}
For small $V$ the fidelity values are close to unity, with the biggest 
deviation observed for the Jastrow-type  wave function. 
 However, at $V \approx 4t$ there is a jump for both TPS wave functions with $D=2$. 
The lower fidelity values can be understood as a consequence of the wave function converging 
to one of  the two symmetry-breaking charge ordered states. 
On the other hand,  the Jastrow-type wave function in the form used in this work cannot break 
the symmetry between the two ordered states, retaining a large overlap with the ED results. 
As for  the $D=3$ TPS wave function, more variational parameters allows for a superposition 
of the two solutions in the range of $V$ considered here, giving a fidelity value close to one.
In Fig.~\ref{fig:bs}(b), we also show the structure factor
$S_q(\pi,\pi)$ and compare it to the ED results.
We notice that all tensor-network based trial states agree with
the exact results very well, while the Jastrow-type wave function
shows the biggest difference near the critical point ($V=2$).

In Fig.~\ref{fig:bs}(c) and (d) a similar analysis is presented for an
$8\times8$ lattice system, for which we compare the results from the
tensor-network projection method to the stochastic series expansion
(SSE) QMC simulations. \cite{SSE1,SSE2,SSE3}
We observe again that the results of the TPS with $D=2$ can be
considerably improved by either increasing the bond dimension to
$D=3$ or adding the Jastrow factors to the TPPS.
The energy errors are very similar for the TPS with $D=3$ and the
TPPS with $D=2$.
The main difference with respect to the SSE method for all input
wave functions we use here is the behavior of these wave functions
around the critical point.
We find that the VMC simulation in the 2D $t$-$V$ model falls into
two minima rather than shows a continuous transition.
The quantitative differences in the structure factor are the smallest for
the $D=3$ TPS, and become more pronounced for $D=2$ TPPS and
 $D=2$ TPS, especially close to $V=2$.

%++++++++++++++++++++++++++++++++++++++++++++++++++++++++++++++++%
\subsection{Spinless fermions}
%++++++++++++++++++++++++++++++++++++++++++++++++++++++++++++++++%

We consider the spinless fermionic $t$-$V$ model at half filling
whose Hamiltonian can be expressed in the same form as that in
Eq.~(\ref{eq:ham}), but now the creation/annihilation operators are fermionic.
It is one of the simplest Hamiltonians for simulating interacting fermions, and it has been
discussed as a model capturing the essential physics of several organic materials 
(at one-quarter filling). \cite{McKenzie01}
The phase diagram of the 2D $t$-$V$ model at half filling was first studied by Gubernatis \textit{et al.} 
using several techniques, including finite-temperature determinantal QMC.\cite{fm_QMC85}
The random-phase approximation calculation of the temperature of the CDW 
transition for small $V$ gives  $T_{CDW} \sim \exp(-\pi/\sqrt{V})$, indicating that the CDW phase is 
the ground state  for all positive values of $V$. 
However, the numerical results of their work do not reach the weak coupling regime.\\
The model has been also considered as a particular case in the field theory study of Hubbard-like lattice models, 
confirming that there is no metallic phase in two dimensions. \cite{Foster}
Recently, a contradictory result has been obtained  by simulating so called string-bond states. \cite{Song13}
The finite size scaling of the structure factor is consistent with a nonzero $V_c$ and the value
estimated for the critical point $V_c$ is $0.45\pm0.02$.

%------------------------------------------------------------------------------------%
\begin{figure}
\begin{center}
\includegraphics[width=8cm]{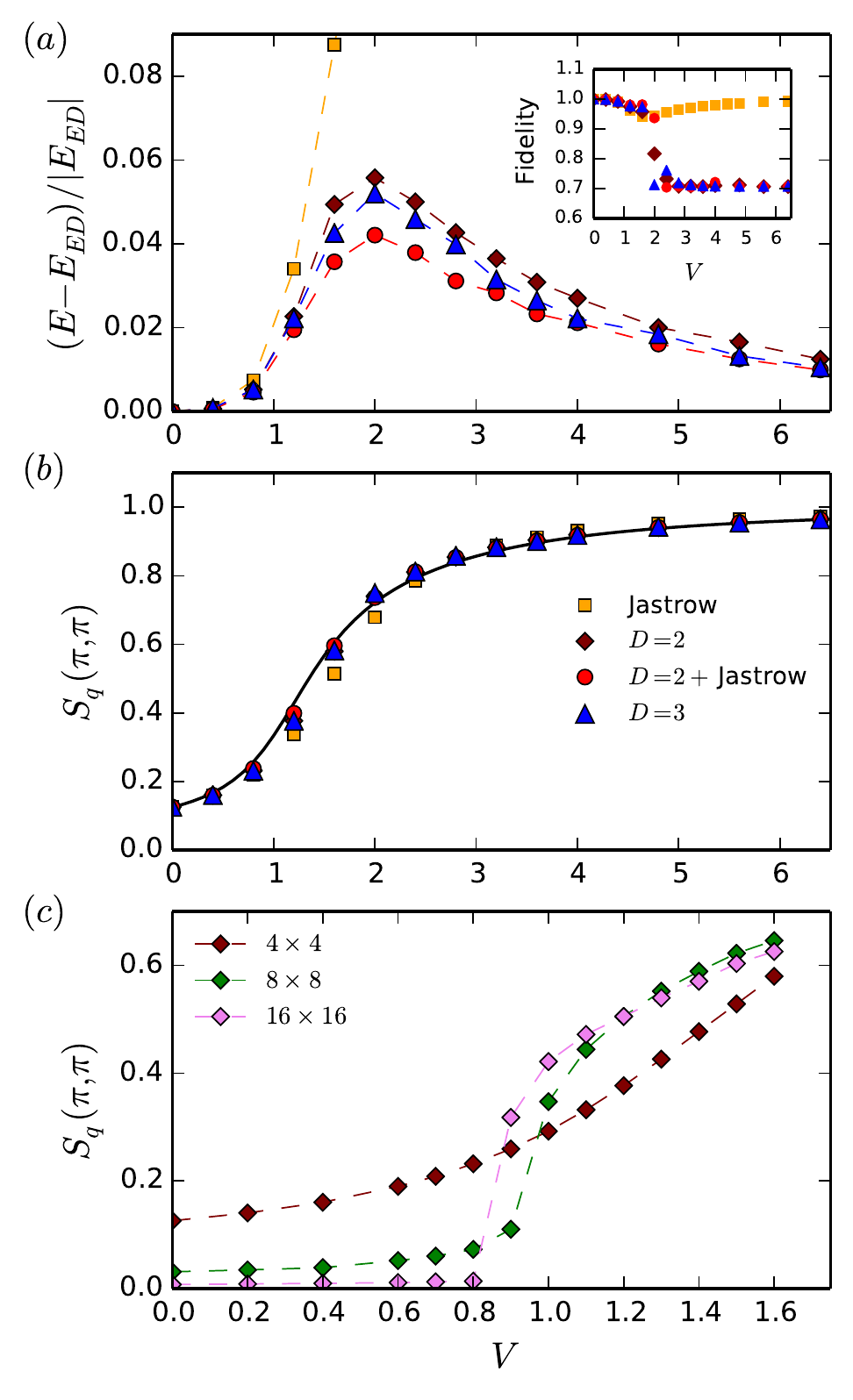}
\end{center}
\caption{Results of VMC simulations of the spinless fermion model.
Panels (a,b) show data for a $4\times4$ lattice system: (a) the relative energy 
error and (inset) the fidelity; (b) the structure factor
$S_q(\pi,\pi)$. All input wave functions include the Slater
determinant: a two-parameter Jastrow-type wave function (orange
squares), a TPPS with the bond dimension $D=2$ (maroon diamonds), a
$D=2$ TPPS with the Jastrow factors (red circles) and a TPPS with
$D=3$ (blue triangles). The ED data for $S_q(\pi,\pi)$ are
denoted by a solid line. (c) The structure factor
$S_q(\pi,\pi)$ for a series of $2^n\times2^n$ clusters ($n=2,3,4$)
calculated by using a TPPS with the bond dimension
$D=2$.  }\label{fig:fm}
\end{figure}
%------------------------------------------------------------------------------------%

In the previous study of the 1D spinless fermion model, using a
matrix-product projected state with a noninteracting fermionic state
as an input wave function, significant improvements to the
ground-state energy and correlations have been found with respect to
regular MPS. \cite{Chou11}
These findings motivate us to propose a TPPS ground state with the
SL as an input wave function, which is the ground state for $V=0$.
It can be easily expressed as a product state in momentum space:
\begin{equation}
|\Psi_{SL}\rangle=\prod_{k<k_{F}}c^{\dagger}_{k}|0\rangle,\label{eq:fl}
\end{equation}
where $k_{F}$ denotes the Fermi momentum.
We apply antiperiodic boundary condition in the $y$ direction in order
to avoid degeneracies at the Fermi level which occur in the case of
periodic boundary conditions for a series of $2^n\times2^n$ clusters
($n=1,2,\ldots$).
Note that it also changes the symmetry of the tensor network:
instead of rotational symmetry used for the bosonic system, we
impose here $x$ and $y$ reflectional symmetry for the elements of
the tensors.\\
We  also use more general input states including a Jastrow-type
wave function,
\begin{equation}
|\Psi_{SL+J}\rangle=\hat{J}_{\gamma,\beta}|\Psi_{SL}\rangle.\label{eq:slater1}
\end{equation}
The Fourier transform of the SL wave function to the site
representation leads to Slater determinants as coefficients in front
of the configuration vectors $|\alpha\rangle$:
\begin{equation}
|\Psi_{SL}\rangle=\sum_{\alpha}\langle\alpha|\Psi_{SL}\rangle|\alpha\rangle.
\label{eq:slater}
\end{equation}
In Fig.~\ref{fig:fm}(a) we present the energy error for SL TPPS for $D=2$ and $3$ 
with ED results ($4\times4$ system).
First of all, the SL TPPS is exact at $V=0$ and remains a good
approximation of the ground-state wave function for small values of $V$.
Unlike in the bosonic case, we do not observe a significant improvement
of the energy with the increasing bond dimension. 
We also consider the $D=2$ TPPS with both Slater determinant and
Jastrow factors.
As mentioned before, adding Jastrow factors practically does not
increase the computational cost, however, the energy error is
reduced with respect to the TPPS using the  Slater
determinant only [cf. red circles and maroon diamonds in
Fig.~\ref{fig:fm}(a)].
The fidelity results shown in the inset to Fig.~\ref{fig:fm}(a) start from unity for $V=0$ and 
then the overlap with the ED wave function slowly decreases as the interaction strength 
increases until $V \approx 2$. 
Above this value all TPPS wave functions fluctuate around 
one of the charge ordered states and  the fidelity values are much lower. 
In Fig.~\ref{fig:fm}(b) we show the structure factor which is in
good agreement with  ED results.\\
In order to study the phase transition to the CDW insulating
phase, we perform the VMC simulations for larger cluster sizes.
We consider the $D=2$ TPPS with Slater determinant
only, and examine the size dependence of the jump of the structure
factor.
In Fig.~\ref{fig:fm}(c), we show the results for the structure
factor $S_q(\pi,\pi)$ computed for a series of clusters
$2^n\times2^n$.
We observe a significant shift with increasing system size, and thus a careful 
finite size scaling analysis is necessary which we defer to a future work.
We note that adding more correlations into the Jastrow type input wave function might
significantly improve the description of the spinless fermion model. \cite{Neu84}
 
%
%++++++++++++++++++++++++++++++++++++++++++++++++++++++++++++++++%
\subsection{Spinful fermions}\label{sec3c}
%++++++++++++++++++++++++++++++++++++++++++++++++++++++++++++++++%

We compare our method to ED data for a spinful fermionic Hubbard model on a $4\times4$ cluster at
half-filling with (anti-)periodic boundary conditions along the $x$
($y$) direction. The fermionic Hubbard Hamiltonian is given by
\begin{eqnarray}
H_{tU}=-t\sum_{\langle
i,j\rangle,\sigma}\left(c^{\dag}_{i\sigma}c_{j\sigma}+H.c.\right)+U\sum_{i}n_{i\uparrow}n_{i\downarrow},\label{e:equ2}
\end{eqnarray}
where $n_{i\sigma}=c^{\dag}_{i\sigma}c_{i\sigma}$ with spin
$\sigma$.

\begin{figure}[t]
\center
\includegraphics[width=0.47\textwidth]{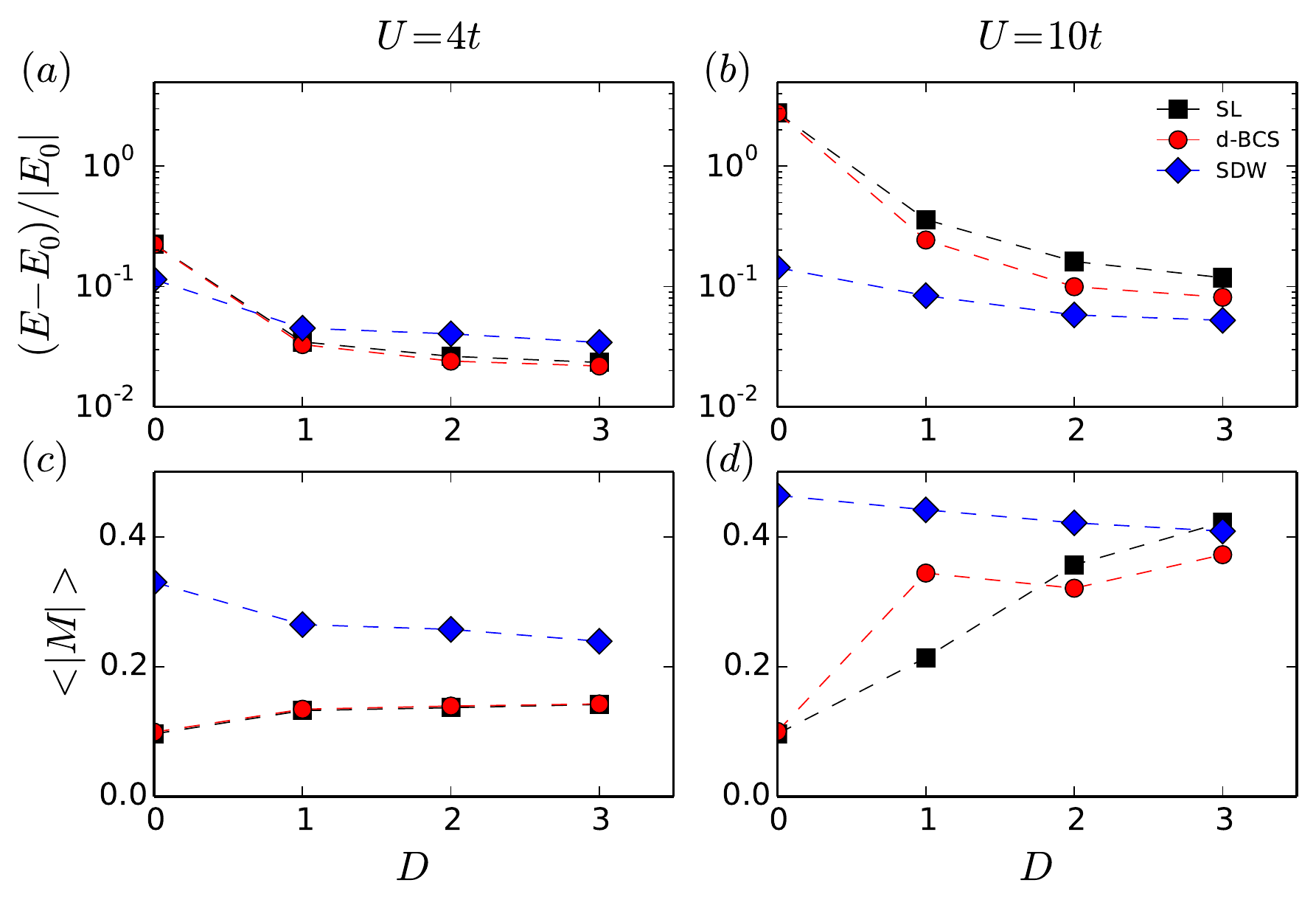}
\caption{Relative energy error of $4\times4$ spinful fermions in the
Hubbard model as a function of the bond dimension $D$ at (a) $U/t=4$
and (b) $U/t=10$ for different input states $|\Psi_{0}\rangle$: SL,
$d$-BCS and SDW. The ED data are taken from
Ref.~\onlinecite{TaharaJPSJ08}. The lower panels show the staggered
magnetization vs $D$ at (c) $U/t=4$ and (d)
$U/t=10$. The $D=0$ values correspond to the unprojected input wave functions.\label{sf_hubbard}}
\end{figure}

We compare a number of different mean field wave
functions as input functions for our method, namely the Slater determinant of spinful fermions (SL), d-BCS and a spin-density wave (SDW).
Details about these standard mean-field wave functions  can be
 found in  Refs.~\onlinecite{GrosAP89,OgataRPP08}.
We first optimize the unprojected wave-functions and then use
the TPPS approach with increasing bond dimension to further optimize the state.
We assume the tensors to be real and without any symmetry.
For our benchmark, we measure the energy and the anti-ferromagnetic (AF) order parameter -- the staggered
magnetization, defined as
\begin{eqnarray}
\langle\left|M\right|\rangle=\left\langle\left|\frac{1}{N}\sum_{i}S_{i}^{z}
e^{i\mathbf{Q}\cdot \mathbf{R}_{i}}\right|\right\rangle, \label{e:equ3}
\end{eqnarray}
 where $\mathbf{Q}=(\pi,\pi)$ is the AF ordering vector, $\mathbf{R}_{i}$ is the position of the fermion at site $i$, and
$S_{i}^{z}=\frac{1}{2}(n_{i\uparrow}-n_{i\downarrow})$.
We show  in
Fig.~\ref{sf_hubbard} that the tensors of the TPPS significantly
improve the variational energy for  all input states $|\Psi_{0}\rangle$.
In the case of the smallest bond dimension $D=1$, the
$T_{[i]}^{j_i}$ are simply scalars for each
$j_i=0,\uparrow,\downarrow,\uparrow\downarrow$ and
thus the resulting projectors resemble onsite Gutzwiller-type projectors.

In the regime of small $U/t$ shown in Fig.~\ref{sf_hubbard}(a)
and (c), the tensor network, acting on these mean-field states
attempt to capture a  weak  staggered magnetization as  $D$ increases.
The SL and BCS states  show slightly lower energies than
the SDW state for any finite $D$.
This result implies that the AF correlation built in the SDW wave
function is  too large and the projectors need to correct for that.
In other words, we have to use a large bond dimension of the tensors
to project out the AF order in the SDW input state.
Choosing a good input wave function thus not only takes care of the
fermionic sign, but it efficiently encodes the correlations between
fermions which would require a  larger bond dimension in regular TPS.

%------------------------------------------------------------------------------------%
%------------------------------------------------------------------------------------%

In the limit of large $U/t$ we expect the ground state to form AF order.
We find in Fig.~\ref{sf_hubbard}(b) indeed that the unprojected
input wave function of the SDW state is much closer to the
ground-state than the unprojected SL and BCS states.
The tensor-product projectors, acting on both SL and BCS input states,
exhibits a staggered magnetization as shown in Fig.~\ref{sf_hubbard}(d).
In particular, already the $D=1$ tensors substantially lower the energy of
the input states.
We furthermore find that the SDW wave function
has the lowest energy for all $D$ as the input
wave-function already takes care of the AF correlation.
In order to approach the same energy level as the SDW state, the
TPPS with SL or BCS state need  larger bond dimensions.
This demonstrates  that TPPS are a promising tool to represent
ground states of 2D Hubbard models.
%
%%%%%%%%%%%%%%%%%%%%%%%%%%%%%%%%%%%%%%%%%%%%%%%%%%%%%
\section{Conclusions}\label{sec4}
%%%%%%%%%%%%%%%%%%%%%%%%%%%%%%%%%%%%%%%%%%%%%%%%%%%%%
We investigated a numerical method based on
tensor-product projected states (TPPS), which combine
recently developed tensor-network based states and
conventional trial wave functions, to study the ground state
properties of quantum many-body systems.
Using conventional VMC methods, we demonstrated
the applicability and flexibility of  TPPS in describing the ground state
energy and correlations for several quantum systems.
Choosing a suitable input state, which captures some known features of the
ground state, we were able to reach significantly lower energy than using
a standard TPS with the  same bond dimension.
Thus the numerical effort to search the exact ground state can be considerably
reduced, both by lowering the cost of tensor contraction and the number
of parameters that need to be optimized.
Our results suggest that the approach would be useful in simulating fermionic
systems where quantum Monte Carlo simulations face the sign problem,
and in the case where the TPS has been used but is not currently capable of
reaching  sufficient accuracies.
Furthermore, a proper quantum-number projection and using symmetric
tensors to filter out excited states with other quantum numbers may potentially
further improve the variational energy of the TPPS.
Finally, we note that TPPS's are capable of expressing states with non-trivial topological
order \cite{chiral} as there are no constraints on the input wave functions.

%%%%%%%%%%%%%%%%%%%%%%%%%%%%%%%%%%%%%%%%%%%%%%%%%%%%%
\section{Acknowledgements}
%%%%%%%%%%%%%%%%%%%%%%%%%%%%%%%%%%%%%%%%%%%%%%%%%%%%%
This work is partly supported by the Chinese Academy of Engineering
Physics and Ministry of Science and Technology (CPC), and the Ministry
of Science and Technology in Taiwan under Grants
No.~100-2112-M-002-013-MY3 (YJK, HWC, OS), 100-2923-M-004-002-MY3 (YJK), and 102-2112-M-002-003-MY3 (YJK). YJK acknowledges  travel support from NCTS in Taiwan. OS gratefully acknowledges support from the visitors program of MPI-PKS Dresden. We also acknowledge the hospitality of the Benasque Center for Science where part of this work was done.


\begin{thebibliography}{30}
\bibitem{XGWenBOOK} X.-G. Wen, \textit{Quantum Field Theory of Many-Body Systems} (Oxford University Press, Oxford, 2004).
\bibitem{Balents10} L. Balents, Nature {\bf 464}, 199 (2010).
\bibitem{WhitePRL92} S. R. White, Phys. Rev. Lett. {\bf 69}, 2863 (1992).
\bibitem{LindenPR92} W. von der Linden, Phys. Rep. {\bf 220}, 53 (1992).
\bibitem{OstlundPRL95} S. \"Ostlund and S. Rommer, Phys. Rev. Lett. {\bf 75}, 3537 (1995).
\bibitem{Verstraete04} F. Verstraete and J. I. Cirac, arXiv:cond-mat/0407066.
\bibitem{VerstraetePRA04} F. Verstraete and J. I. Cirac, Phys. Rev. A {\bf 70}, 060302 (2004).
\bibitem{ShiPRA06} Y.-Y. Shi, L.-M. Duan, and G. Vidal, Phys. Rev. A {\bf 74}, 022320 (2006).
\bibitem{SchuchPRL08} N. Schuch, M. M. Wolf, F. Verstraete, and J. I. Cirac, Phys. Rev. Lett. {\bf 100}, 040501 (2008).
\bibitem{SandvikPRL08} A. W. Sandvik, Phys. Rev. Lett. {\bf 101}, 140603 (2008).
\bibitem{MurgPRB10} V. Murg, F. Verstraete, \"O. Legeza, and R. M. Noack, Phys. Rev. B {\bf 82}, 205105 (2010).
\bibitem{MezzacapoNJP09} F. Mezzacapo, N. Schuch, M. Boninsegni, and J. I. Cirac, New J. Phys. {\bf 11}, 083026 (2009).
\bibitem{ChanglaniPRB09} H. J. Changlani, J. M. Kinder, C. J. Umrigar, and G. K.-L. Chan, Phys. Rev. B {\bf 80}, 245116 (2009).
\bibitem{Hubener09-11} R. H\"ubener \textit{et al.}, Phys. Rev. A {\bf 79}, 022317 (2009); R. H\"ubener \textit{et al.}, Phys. Rev. B {\bf 84}, 125103 (2011).
\bibitem{HubenerNJP10} R. H\"ubener, V. Nebendahl, and W. D\"ur, New J. Phys. {\bf 12}, 025004 (2010).
\bibitem{MurgPRA07} V. Murg, F. Verstraete, and J. I. Cirac, Phys. Rev. A {\bf 75}, 033605 (2007).
\bibitem{MurgPRB09} V. Murg, F. Verstraete, and J. I. Cirac, Phys. Rev. B {\bf 79}, 195119 (2009).
\bibitem{BauerJSM09} B. Bauer, G. Vidal, and M. Troyer, J. Stat. Mech.: Theory Exp. P09006 (2009).
\bibitem{Wang11} L.Wang, I. Pi\v zorn, and F. Verstraete, Phys. Rev. B {\bf 83}, 134421 (2011).
\bibitem{VidalPRL07} G. Vidal, Phys. Rev. Lett. {\bf 99}, 220405 (2007).
\bibitem{GuPRB08} Z.-C. Gu, M. Levin, and X.-G. Wen, Phys. Rev. B {\bf 78}, 205116 (2008).
\bibitem{JiangPRL08} H. C. Jiang, Z. Y. Weng, and T. Xiang, Phys. Rev. Lett. {\bf 101}, 090603 (2008).
\bibitem{YJK11-12} L. Wang, Y.-J. Kao, and A. W. Sandvik, Phys. Rev. E {\bf 83}, 056703 (2011); J.-F. Yu and Y.-J. Kao, Phys. Rev. B {\bf 85}, 094407 (2012).
\bibitem{GuPRB13} Z.-C. Gu, Phys. Rev. B {\bf 88}, 115139 (2013).
\bibitem{Wouters14} S. Wouters, B. Verstichel, D. Van Neck, and G. K.-L. Chan, Phys. Rev. B {\bf 90}, 045104 (2014).
\bibitem{Clark14} B. K. Clark and H. J. Changlani, arXiv:1404.2296.
\bibitem{Neu84} E. Neuscamman H. Changlani, J. Kinder, and G. K.-L. Chan, Phys. Rev. B {\bf 84}, 205132 (2011).
\bibitem{Neu85} E. Neuscamman , C. J. Umrigar, and G. K.-L. Chan, Phys. Rev. B {\bf 85}, 045103 (2012).
\bibitem{LegPRB03} \"O. Legeza and J. S\'olyom Phys. Rev. B {\bf 68}, 195116 (2003).
\bibitem{LegPRA11} G. Barcza, \"O. Legeza, K. H. Marti and M. Reiher, Phys. Rev. A {\bf 83}, 012508 (2011).
\bibitem{Chou11} C.-P. Chou, F. Pollmann and T.-K. Lee, Phys. Rev. B {\bf 86}, 041105(R) (2012).
\bibitem{Sorella01} S. Sorella, Phys. Rev. B {\bf 64}, 024512 (2001).
\bibitem{YunokiPRB06} S. Yunoki and S. Sorella, Phys. Rev. B {\bf 74}, 014408 (2006).
\bibitem{Capello06} M. Capello, \textit{Variational description of Mott insulators}, PhD thesis, SISSA (2006).
\bibitem{Hebert01} F. H\'ebert, G.G. Batrouni, R.T. Scalettar, G. Schmid, M. Troyer and A. Dorneich, Phys. Rev. B {\bf 65}, 014513 (2001).
\bibitem{SSE1} A. W. Sandvik and J. Kurkij\"arvi, Phys. Rev. B {\bf 43}, 5950 (1991).
\bibitem{SSE2} A. W. Sandvik Phys. Rev. B {\bf 59}, R14157(R) (1999).
\bibitem{SSE3} O. F. Sylju\aa sen and A. W. Sandvik  Phys. Rev. E {\bf 66}, 046701 (2002).
\bibitem{McKenzie01} R. H. McKenzie, J. Merino, J. B. Marston, and O. P. Sushkov, Phys. Rev. B {\bf 64}, 085109 (2001).
\bibitem{fm_QMC85} J. E. Gubernatis, D. J. Scalapino, R. L. Sugar, and W. D.Toussaint, Phys. Rev. B {\bf 32}, 103 (1985).
\bibitem{Foster} M. S. Foster and A. W. W. Ludwig, Phys. Rev. B {\bf 77}, 165108 (2008)
\bibitem{Song13} J. -P. Song and R.T. Clay, Phys. Rev. B {\bf 89}, 075101 (2014).
\bibitem{GrosAP89} C. Gros, Ann. Phys. {\bf 189}, 53 (1989).
\bibitem{OgataRPP08} M. Ogata and H. Fukuyama, Rep. Prog. Phys. {\bf 71}, 036501 (2008).
\bibitem{TaharaJPSJ08} D. Tahara and M. Imada, J. Phys. Soc. Jpn. {\bf77}, 114701 (2008).
\bibitem{chiral} T. B. Wahl, H. H. Tu, N. Schuch, and J. Cirac, Phys. Rev. Lett., 111, 236805 (2013)
\end{thebibliography}
\end{document}